\documentclass[letterpaper, 10 pt, conference]{ieeeconf}  % Comment this line out if you need a4paper
\IEEEoverridecommandlockouts                              % This command is only needed if 
\usepackage{graphics} % for pdf, bitmapped graphics files
\usepackage{epsfig} % for postscript graphics files
\usepackage{mathptmx} % assumes new font selection scheme installed
\usepackage{times} % assumes new font selection scheme installed
\usepackage{amsmath} % assumes amsmath package installed
\usepackage{amssymb}  % assumes amsmath package installed
\usepackage{multirow}
\usepackage[table]{xcolor}

\definecolor{mygray}{gray}{0.9}

\title{\LARGE \bf
Deep Learning based acoustic measurement approach for robotic applications on orthopedics
}

\author{Bangyu Lan$^{1}$, Momen Abayazid$^{1}$, Nico Verdonschot$^{1,2}$, Stefano Stramigioli$^{1}$ and Kenan Niu$^{1}$
    \thanks{$^{1}$Robotics and Mechatronics, University of Twente, Enschede, AE, The Netherlands, $^{2}$Orthopaedic Research Lab, Radboud University Medical Center, Nijmegen, the Netherlands}}

\begin{document}
\maketitle
\thispagestyle{empty}
\pagestyle{empty}

%%%%%%%%%%%%%%%%%%%%%%%%%%%%%%%%%%%%%%%%%%%%%%%%%%%%%%%%%%%%%%%%%%%%%%%%%%%%%%%%
\begin{abstract}
In Total Knee Replacement Arthroplasty (TKA), surgical robotics can provide image-guided navigation to fit implants with high precision. Its tracking approach highly relies on inserting bone pins into the bones tracked by the optical tracking system. This is normally done by invasive, radiative manners (implantable markers and CT scans), which introduce unnecessary trauma and prolong the preparation time for patients. To tackle this issue, ultrasound-based bone tracking could offer an alternative. In this study, we proposed a novel deep-learning structure to improve the accuracy of bone tracking by an A-mode ultrasound (US). We first obtained a set of ultrasound dataset from the cadaver experiment, where the ground truth locations of bones were calculated using bone pins. These data were used to train the proposed CasAtt-UNet to predict bone location automatically and robustly. The ground truth bone locations and those locations of US were recorded simultaneously. Therefore, we could label bone peaks in the raw US signals. As a result, our method achieved sub-millimeter precision across all eight bone areas with the only exception of one channel in the ankle. This method enables the robust measurement of lower extremity bone positions from 1D raw ultrasound signals. It shows great potential to apply A-mode ultrasound in orthopedic surgery from safe, convenient, and efficient perspectives.
\end{abstract}

%%%%%%%%%%%%%%%%%%%%%%%%%%%%%%%%%%%%%%%%%%%%%%%%%%%%%%%%%%%%%%%%%%%%%%%%%%%%%%%%
\section{INTRODUCTION}
Measuring bone position is important to understand the positions and kinematics of the lower extremity, for example, in robotic total knee replacement arthroplasty \cite{li2022accuracies} and wearable exoskeleton \cite{meier2023evaluating}. However, traditional measurements, such as CT scan, skin markers, or implantable markers, unexpectedly introduce accumulated distance errors, unnecessary trauma, radiation risks, and infections.

To precisely measure distance without unnecessary trauma, previous studies used data from multiple sensors. \cite{fang2022second} worked on measuring and evaluating fingertip distances using optical sensors and ultrasound probes. \cite{shi2022novel} regarded that merely using ultrasound was difficult to predict movements of the lower extremities; instead, they used data from both EEG and sEMG recordings. A-mode ultrasound has recently been proposed for bone model reconstruction \cite{chen2020reconstruction, gebhardt2023femur}, registration and surgery robotics \cite{guinebretiere2020feasability, zhang2021hybrid, liu2023accurate}, as it is easy to deploy and feasible to collect data from different bone locations via multiple channels simultaneously. Reconstruction and registration of bony surface can be carried out using these data in different bone locations. 

For example, ultrasound was studied to apply in lower extremity motion tracking and bone measurement for surgical robotics, which was a novel convenient and noninvasive method \cite{niu2018novel, niu2018feasibility, niu2018situ, niu2018measuring}. However, in these approaches, distance measurement was mostly based on expert knowledge, by knowing the approximate range of bone peak's locations and the general shapes of the peaks, which purely based on experience and was lacked of robustness and generalization, as a little disturbance brought by processing or measurement noise could easily change profiles of the peak, making it difficult to identify. In our approach, we used the generalized and adaptability of deep learning to automatically measure bone reflection peaks in US signals without additional knowledge, making ultrasound-based measurements more applicable and automatic. In addition, it helped to ease the difficulty of deployment in total knee replacement arthroplasty and other surgical robot applications.

In previous studies, to analyze 1D medical signals using deep learning, the U-Net structure was used to recognize and identify ECG peaks to diagnose heart disease \cite{moskalenko2020deep} by its contextual information preservation and feature localization capability. The different resolution perceptual fields can capture different sizes of peak profiles. However, it was difficult to recognize the various reflection peaks in the US, as the reflection peaks are more sparse, random, and have diverse profiles in different channels. To solve the issue, we exploited UNet's localization pattern by using a cascade U-Net structure for different perception resolutions, connected with a novel sampling-based proposal mechanism. In addition, an attention framework was introduced to filter out features that are irrelevant to the target peak range \cite{chen2022multiple}. This helped to continually improve the perception of the peak profile. Through these designs, our CasAtt-UNet (\textbf{Cas}caded \textbf{Att}ention \textbf{UNet}) could easily recognize the sparse and effective peak profiles in the US signals with high accuracy. In addition, since the proposed CasAtt-UNet could infer signal peaks and calculate bone location efficiently, it could achieve real-time bone measurement in the surgical robot application, e.g. total knee replacement arthroplasty.

To the best of our knowledge, few studies focused on analyzing and interpreting the 1D ultrasound signal to detect bones using deep learning method. In this paper, we aim to introduce a deep learning-based 1D raw signal peak locating approach for accurate bone position measurement. The proposed method showed great potential for using deep learning to analyze complex and random ultrasound signals. For future development, this can be deployed on surgical robot arms to guide precise total knee replacement arthroplasty. It is worth noting that the proposed technique could also hold the potential for adaptation to other robotics surgery where bone motion tracking is essential. This is also beneficial to improve the accuracy of bone kinematics measurement. 

\begin{figure}[t!]
\centering
\includegraphics[width=1.0\linewidth]{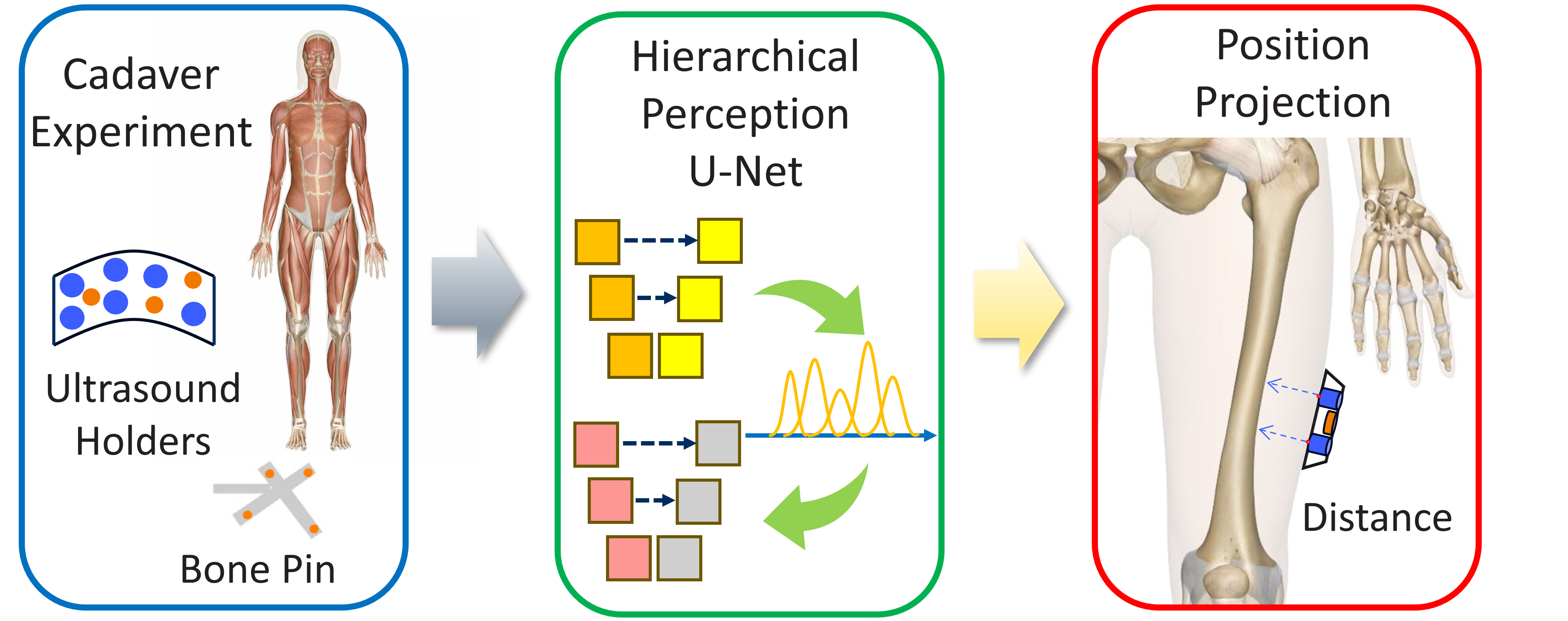}
\caption{Pipeline overview: Our method had three steps (from left to right): performed cadaver experiment to collect ultrasound signals \textbf{(network input)} and calculate bone positions \textbf{(network output)}, Use the dataset to train CasAtt-UNet, recover the bone position and evaluate.}
\label{pipeline}
\end{figure}

\section{MATERIAL AND METHOD}
Our method contained three parts in Fig. \ref{pipeline}: Data collection from a cadaver experiment, CasAtt-UNet training, and validation of the inference result. The US data and positions of the optical markers were collected from a full-body cadaver specimen, where bone pins were inserted into the femur and tibia for the reference locations of the bones. 

Firstly, ultrasound holders were attached to the cadaver leg and tracked using the 3D optical tracking system. Each holder has multiple US transducers to collect US reflection waves in one anatomical locations on the femur or tibia. Thanks to the bone pins, the ground truth position of the two bones could be recalculated, and the reference positions of the attached US holders were also recorded concurrently. The precise bones locations were then calculated with respect to the positions of the attached US holders (i.e., that of each ultrasound probe). Subsequently, the reference distances between each ultrasound transducer and the underlying bone surface could be derived. Because bone locations and bone reflection peaks in the raw 1D ultrasound signal are correlated, the ultrasound signal with bone locations was used to train our CasAtt-UNet. In the end, the precision at different bone locations was evaluated, and the performance of our model was reported.

\begin{figure}[t!]
\centering
\includegraphics[width=1.0\linewidth]{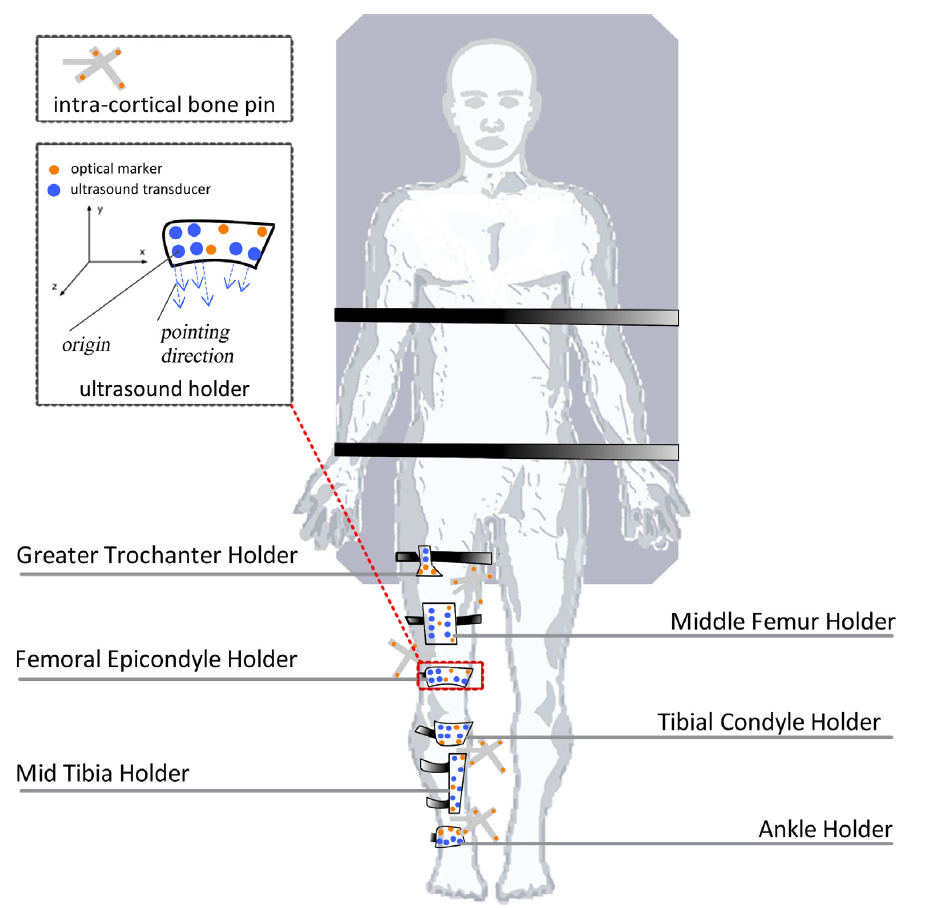}
\caption{Location of the US holders\cite{niu2018situ}: Our ultrasound holders were installed on the six locations of the left leg: Trochanter, Mid Tibia, Femur Epicondyle, Tibia Epicondyle, Mid Tibia and Ankle. Each holder was tied using bandages. Notice that the distribution of optical markers and transducers were different with the ones in the image.}
\label{cadaver_experiment}
\end{figure}

\subsection*{A. Experimental Setup and Data Acquisition}
To ensure the accuracy and functioning of our method, a human cadaver specimen (male, 79kg, 179cm) was used to acquire dataset. This has been approved by Radboud University Medical Center (Radboud UMC), Nijmegen, the Netherlands. A full leg (from pelvic to foot) was CT scanned (TOSHIBA Aquilion ONE, voxel size of $0.755 mm \times 0.755 mm \times 0.500 mm$). Subsequently, 3D geometric models of the femur and tibia were segmented using Mimics 17.0 (Materialise N.V., Leuven, Belgium). 

This dataset was collected in our previous study \cite{niu2018situ}. During the experiment, each ultrasound holder contained three LED optical markers and several 7.5MHz A-mode ultrasound transducers (Imasonic SAS, Vorayl'Ognon, France), which were used to acquire ultrasound echos. There were in total 30 A-mode ultrasound transducers and 18 LED optical markers distributed on the 6 ultrasound holders (holders were designed from CAD models). Also there were 16 LED optical markers distributed on the four bone pins. The sample rate of the entire US tracking system was 20 Hz. An optical tracking system (Visualeyez VZ4000v trackers, PTI Phoenix Technologies Inc., Vancouver, Canada) was operated at 100 Hz to track the 3D locations of the US probes. The ultrasound signal was acquired and synchronized with the optical tracking system in the Diagnostic Sonar FI Toolbox (Diagnostic Sonar Ltd., Livingston, Scotland). The origin and direction of the ultrasound beam were determined from the calibration method \cite{niu2018feasibility}. 

To record the location of the bones, four bone pins were inserted into different parts of the femur and tibia, and six ultrasound holders were fixed at different locations of the femur and tibia. Throughout the experiment to collect the dataset, the leg was actively maneuvered through a cyclic flexion and extension process to emulate the swing phase of the gait cycle, which simulated the condition of a person's walking. This caused changes in distance between bones and transducers, resulting in changes of peak locations in ultrasound signals.

After data collection, we gathered US signals and trajectories of the attached optical markers from the holders. These data were used to reconstruct the actual locations of US holders above bones and the directions of US waves. Consequently, the actual bone depth can be derived using the bone location, the origin of US waveform, and the US wave directions. 

In total, a dataset was acquired that contained 1017 continuous samples (moments) recorded from leg movement. In each sample, there were ultrasound echos from 30 transducer channels and 3D positions of 34 optical markers (16 bone pin markers and 18 US holder markers). We noticed that for most bone peaks in the US signals from Trochanter and Mid Tibia, they have been attenuated or disappeared. The missing number had already exceeded half of all acquired samples, which was clearly not suitable for network training. Therefore, the dataset in these two locations was directly discarded. The rest dataset was checked and screened too. Finally, the 1D ultrasound signals in four anatomical areas (Fig. \ref{cadaver_experiment}) were collected: Femur Epicondyle, Tibia Epicondyle, Mid Tibia and Ankle. The holder at each location contained three optical markers and several transducers. Totally nine channels in the four anatomical locations with the trajectories of twelve LED optical markers were suitable for training the CasAtt-UNet.

\begin{figure}[t!]
\centering
\includegraphics[width=1.0\linewidth]{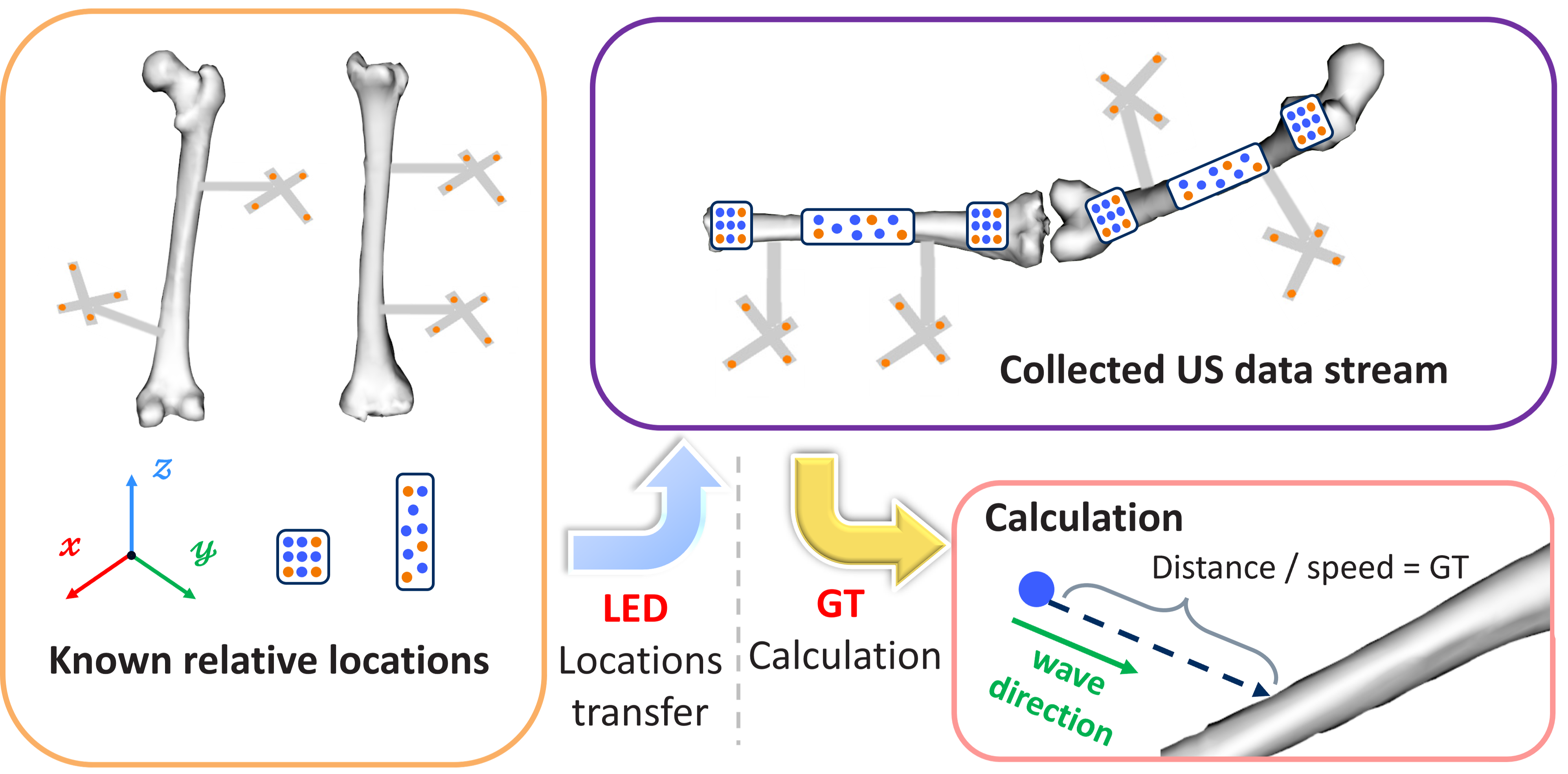}
\caption{Steps to determine ground truth labels: The optical markers in the predefined and the experiment case determined the transformation, which was used to transform US transducers and waves directions to the experiment coordinate frame. The label was calculated using Euclidean distance and the speed of US.}
\label{transformed_position}
\end{figure}

\subsection*{B. Bone Location Calculation and US Signal Peak Labeling}
After cadaver experiment, we labeled the ground-truth position of bone peaks in the 1D ultrasound signals. What we had were the following: 1, 3D geometric surface of femur \& tibia and 3D positions of bone pins; 2, 3D positions and distributions of the US transducers and optical markers in the holders; 3, 3D positions of optical markers (bone pins and holders) in the experiment. The processing procedure was to transfer transducers, femur and tibia from the predefined coordinate frame (e.g. CT frame and CAD frames) to the experiment coordinate frame (i.e. the coordinate frame of the cadaver specimen in the experiment) in each moment. The process was shown in Fig. \ref{transformed_position}. Firstly the transformation $^{R}_{H}{T}$ that align optical markers $\{H\}$ in the predefined model with the ones in the experiment $\{R\}$ was calculated, which was $f(\{H\}, \{R\})$. Then this matrix $^{R}_{H}{T}$ was used to transfer US transducers positions $\{P_H\}$ to the ones in the experiment $\{P_R\}$, which was $^{R}_{H}{T} \cdot p_H$. The calculation also kept the wave directions unchanged. The same process was done for both femur and tibia bones, where 16 optical markers on four bone pins were used to transform. The calculation was as following. This transformation process applied to all US transducers positions.

\begin{align}\label{transfrom}
^{R}_{H}{T} &= f(\{H\}, \{R\}) \\
p_R &= ^{R}_{H}{T} \cdot p_H
\end{align}

\begin{figure}[t!]
\centering
\includegraphics[width=1.0\linewidth]{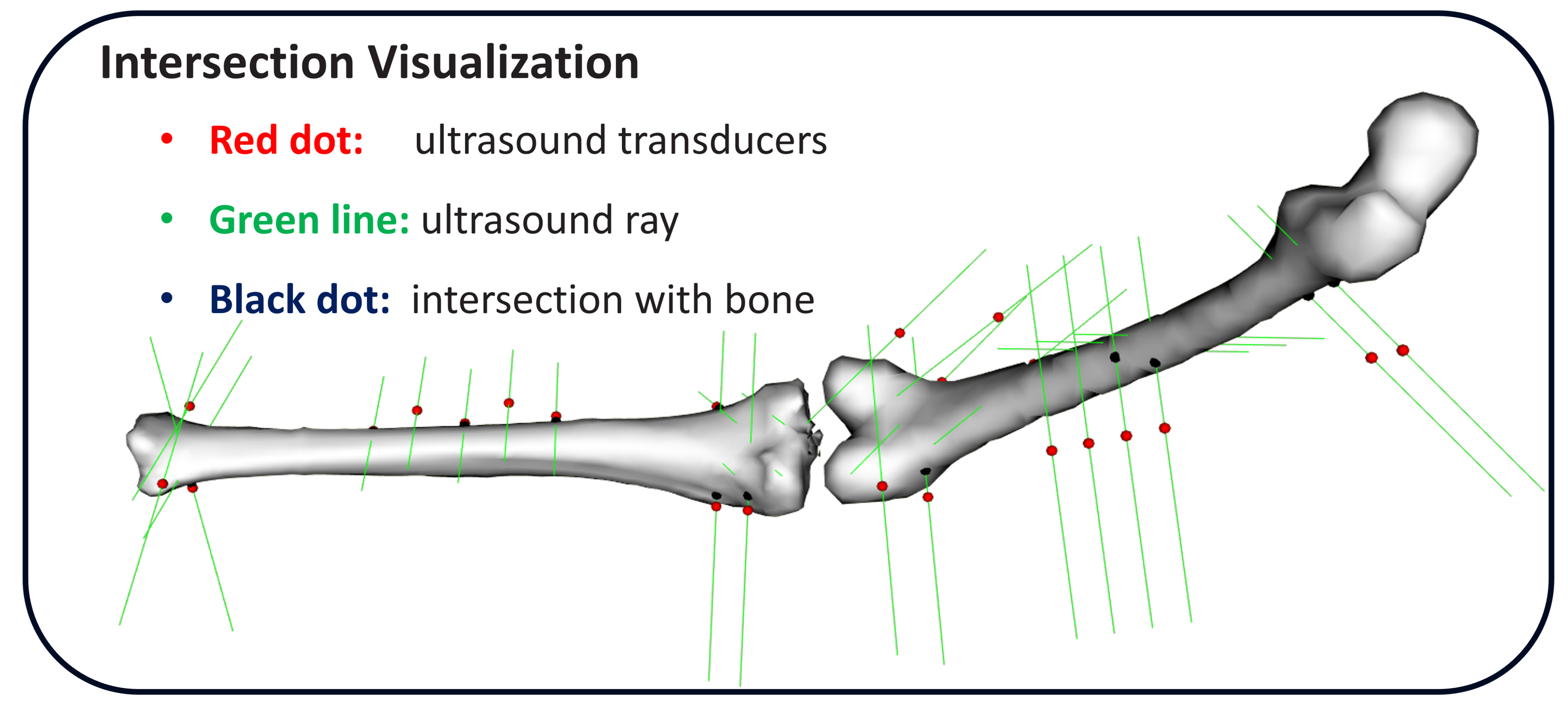}
\caption{Intersection between bones and US waves: This showed one moment in the experiment that the US waves intersected with the bony surface. The intersection positions produced ground truth locations. Red dots were transducer probes positions. Black dots were intersection positions. Green lines were waves directions. The ground truth distance (later used for labeling) was the line segments between black dots and red dots. For some waves there was no black dot as there was no intersection.}
\label{gt_visualize}
\end{figure}

After obtaining all relative positions and wave directions, we could render the US waves starting from the transducers and ending on the bony surface. The distance between intersection positions and the transducer probes were calculated and shown in Fig. \ref{gt_visualize}. After knowing the 3D positions of transducer probes and the intersections, we could calculate the Euclidean distance $d (mm)$ between them (Green line segments cropped by red and black dots). To annotate the corresponding reflected peak position in 1D ultrasound signals, the bone peak location was calculated using Equation (\ref{gt_location}), represented as the index of units ($idx$). Here we assumed the ultrasound speed under skin was $v = 1540m/s$ from \cite{azhari2010appendix}, the sampling rate was $f_s = 40$MS.

\begin{equation}\label{gt_location}
idx = \frac{d}{(2 \times \frac{v}{f_s} \times 1000)}
\end{equation}
, where $2 \times \frac{v}{f_s} \times 1000$ was the one unit length along the 6760 length of the 1D raw ultrasound signal. From our calculation, one unit length in the signal is equal to 0.01925 $mm$ in the distance measurement, which is the minimum accuracy level.

After getting the depth point in the signal, the location in the 1D signal could be annotated as the training label. Using the raw US signals and labels, we could train our CasAtt-UNet.

\subsection*{D. Overview of Cascade Attention U-Net}
\begin{figure}[t!]
\centering
\includegraphics[width=1.0\linewidth]{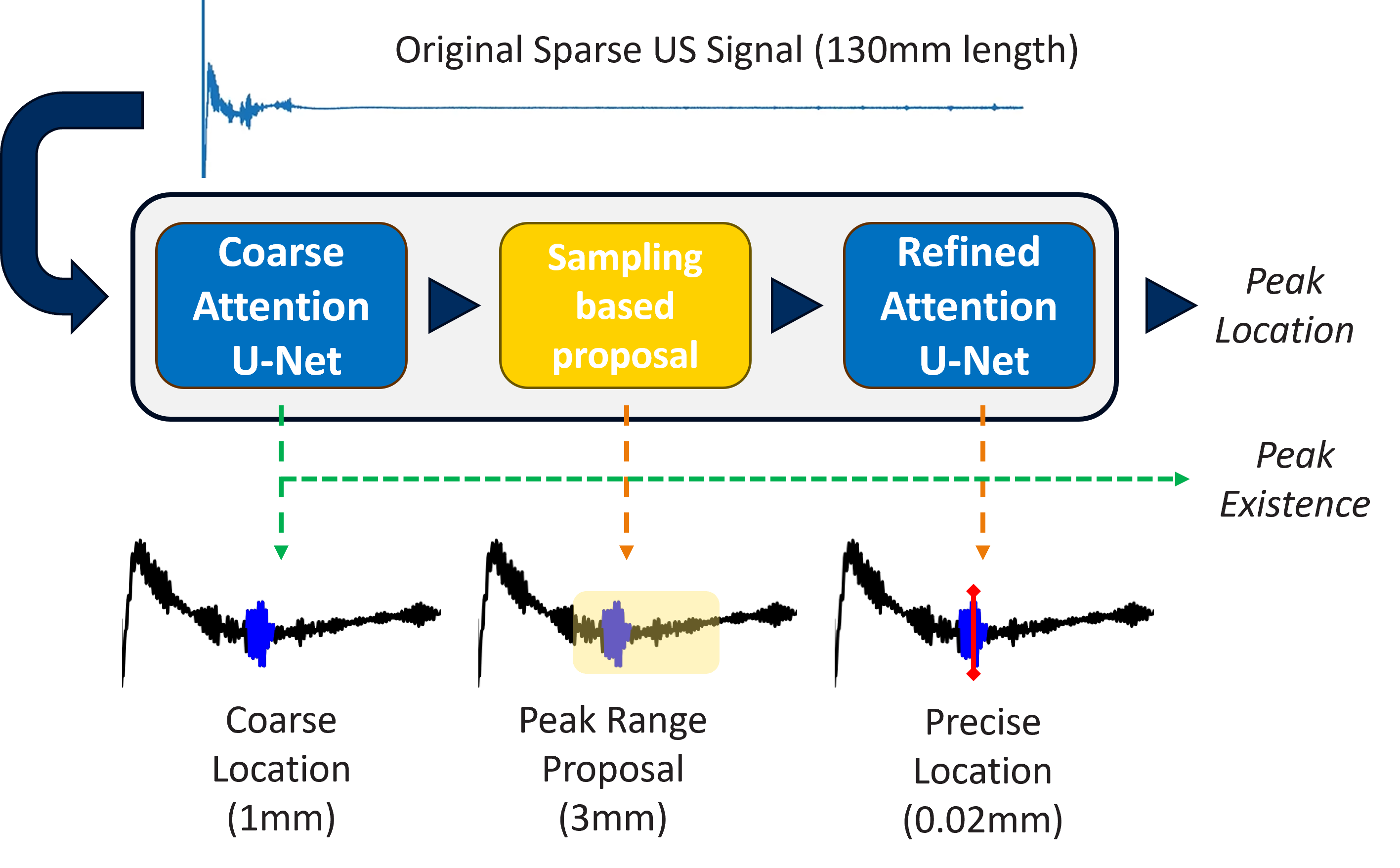}
\caption{Pipeline of Algorithm: Given a 130mm US raw signal, the first coarse U-Net captured the approximate range (1mm) of bone peak. The output was used to segment a continuous length (3mm) of signal region, which is the input of the refined U-Net to determine the exact position of the bone peak.}
\label{algorithm}
\end{figure}

The proposed CasAtt-Unet was shown in Fig. \ref{algorithm}. It composed of coarse attention U-Net, sampling-based proposal, and refined attention U-Net. As the raw US signal occupied 130$mm$ while the original peak annotation was only one index interval (1 index = 0.01925$mm$), a hierarchical structure was required for narrowing down the range. To do this, a coarse attention U-Net was first used to determine the existence of bone peak and capture the approximate range (1mm). From this region, a novel sampling-based mechanism proposed the most likely region of the bone peak. Based on the proposal, the refined U-Net predicted the exact peak location.

\subsection*{E. Structure of Attention U-Net}
\begin{figure}[t!]
\centering
\includegraphics[width=1.0\linewidth]{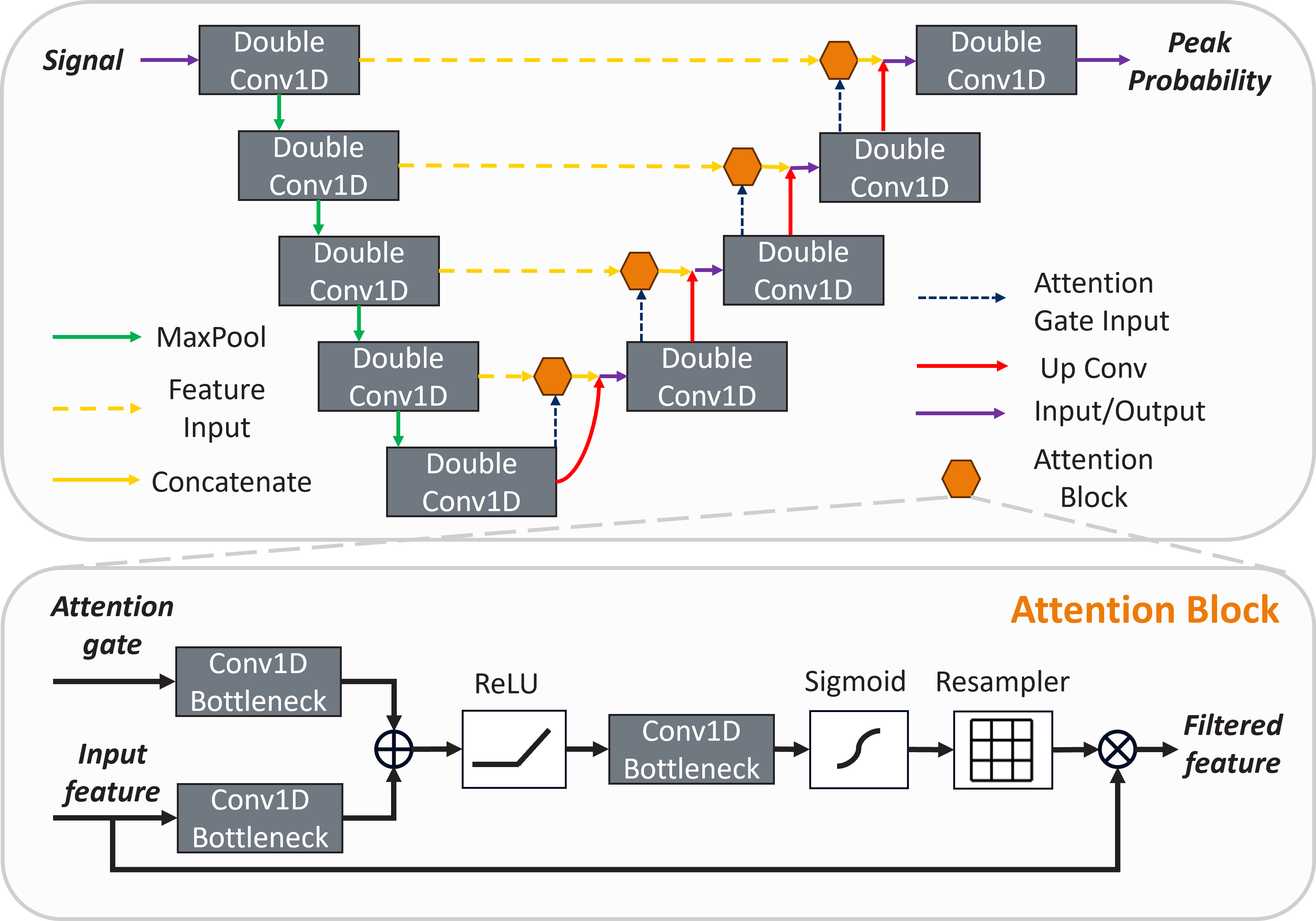}
\caption{Attention U-Net Structure: We replaced the 2D convolution in normal U-Net to 1D convolution, and added additional attention block to increase the perception ability. In each orange hexagon connection (attention block), the attention signal was from the deeper layer, while the input feature was from the left side of U-Net. The output was concatenated with the up convolution result of the deeper layer and input to the double conv1D block.}
\label{att-UNet}
\end{figure}

The proposed coarse and refined attention U-Net was shown in Fig. \ref{att-UNet}, which was inspired from \cite{moskalenko2020deep}. Compared with the normal U-Net in image domain, the convolution kernels had been replaced from two dimension to one dimension. In addition, an attention block similar to \cite{oktay2018attention} was inserted in every skip connection between two sides of U-Net. The output from the left side worked as the input feature of the attention block, filtered by the attention signal from the deeper layers, as the features from deeper layers contained abstract and general space knowledge that can filter out large irrelevant areas in the signal. In this way, the CasAtt-Unet performed more robustly when dealing with more complex and random signals. During training, because the peak summit was only one unit length (0.01925 $mm$) along the whole range, to facilitate network training, the range of peak was increased to 5 units (about 0.1 $mm$) for the Refined location. For the coarse UNet, the coarse range has been increased to 50 units ( approx. 1 $mm$). The network output had the whole signal length consisting of bone peak probability in each unit. This would be threshold by 0.5 probability to the bone peak segment. To recover the exact peak position (i.e., bone depth), we used the middle point of this segment to calculate the depth distance by multiplying the middle point unit index number with the unit interval length.

\subsection*{F. Mechanism of Sampling-based Proposal}
\begin{figure}[t!]
\centering
\includegraphics[width=1.0\linewidth]{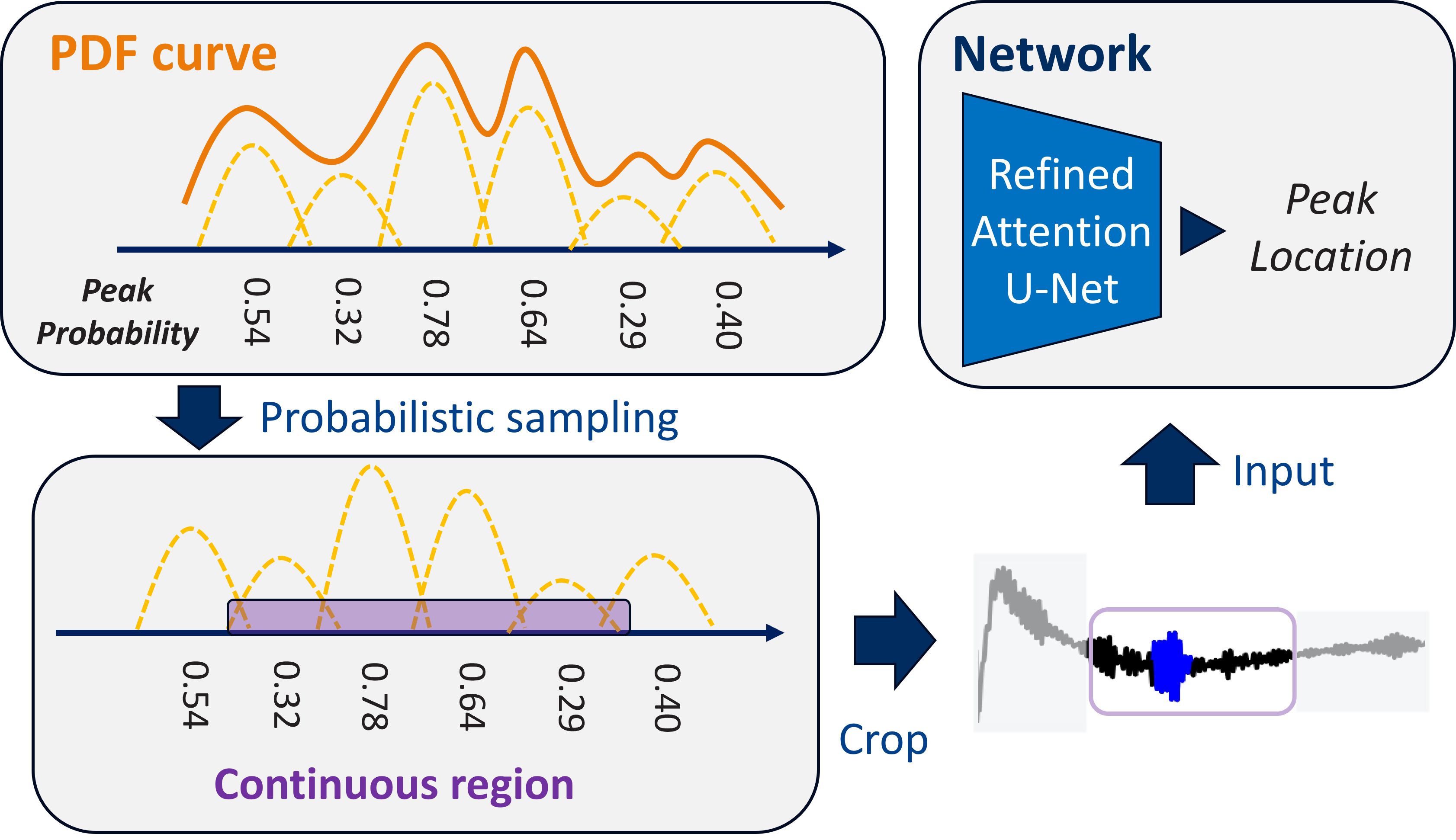}
\caption{Sampling-based Proposal: This structure connected the coarse and refined attention U-Net. The Gaussian distribution was built on each unit position of the prediction. Then a mixture PDF curve was built, on which a continuous region was sampled and cropped the original raw signal. The result was the input of the refined U-Net.}
\label{sampling_proposal}
\end{figure}

To have a candidate region from the coarse U-Net output, a sampling-based proposal method was required and shown in Fig. \ref{sampling_proposal}. A standard Gaussian distribution (mean=1.0, std=1.0) curve, whose amplitude was determined by the predicted probability, was built on each unit of the whole signal. They were combined and summed to have a mixture possibility density curve. Based on this PDF, a continuous fixed-length region was probabilistic sampled. The start and end location cropped the original US signal to be the input of refined attention U-Net, which continuously identified the bone peak in this local region. The benefit of probabilistic approach instead of the deterministic one was that: Each time the refined U-Net can be confronted with a random region even if the coarse U-Net output was the same, the refined U-Net was forced to learn the meaningful peak profile instead of memorizing peak position. This enabled CasAtt-UNet's robust detection when only trained using a limited dataset. During inference, instead of sampling from the PDF, the peak position was directly decided by the largest probability position in this region.

\subsection*{G. Training and Evaluation}
To train a dataset with highly unbalanced foreground and background labels, we introduced dice loss and cross-entropy loss simultaneously. The dice loss was written in Equation (\ref{dice_loss}). It is defined as one minus dice coefficient, which is widely used in medical image segmentation: The numerator is defined as twice of intersection between ground truth and prediction, while the denominator is defined as the total probability of prediction and ground truth. The $\epsilon$ can stabilize the training and prevent zero division. For cross-entropy loss, we constructed one-hot vectors for each unit prediction, then did a binary class softmax operation on the U-Net output. The loss was calculated between onehot labels and softmax results. These two losses were used for both the coarse and refined U-Nets' training. 

\begin{equation}\label{dice_loss}
Dice Loss = 1-\frac{2\sum_{i=0}^n(p_i^{pred}*p_i^{true})+\epsilon}{\sum_{i=0}^np_i^{pred}+\sum_{i=0}^np_i^{true}+\epsilon}
\end{equation}

We split the whole dataset into 8:2 for training and testing. For network training, we augmented for 10 times by shifting the signal x-axis up to 1000 units. To train the model, the first coarse U-Net was trained for first 30 epochs so that a relatively accurate coarse region could be generated, then the second refined U-Net was trained for another 20 epochs using the output of Sampling-based Proposal to get the exact positions of bone peaks. 

For evaluation, the inference result from the refined UNet could be represented as a sequence of bone peak probability: $S_{prob} = \{p_1,p_2,...,p_m\}$. After filtering the outputs using the softmax and 0.5 threshold probability, we had a feasible segment of indices in the signal to represent peak positions: $S_{index}\{idx_1, idx_2,..., idx_n\}$. The exact peak prediction was defined as the middle of this segment, which was expressed as $(max(S_{index})+min(S_{index}))/2$. With the unit index length, the bone peak depth prediction could be easily calculated by multiplying the position unit index with the interval length from Equation (\ref{gt_location}): $2 \times \frac{v}{f_s} \times 1000$. The whole step was in Equation (\ref{predicted_location}).

\begin{equation}\label{predicted_location}
Position = Interval \times (max(S_{index})+min(S_{index}))/2
\end{equation}

With the prediction and the ground truth bone depth, our method performance could be evaluated by counting the percentage when the bias of ground truth and prediction was lower than 0.5mm (ACCURACY). We also calculated the average bias (BIAS) and the detection rate whether the bone peak existed or not (FIND PEAK\%). We reported the results in the following sections.

\section{EXPERIMENTAL RESULTS}
Here the method's performance was compared with the traditional method. For each body area (the US holder location), we selected 2 to 3 typical channels, in which the peaks were clear to see and the number of peaks was over half of total 1017 samples. This allowed for the training and facilitated the learning process. 

\subsection*{A. Traditional Method Result}
For the traditional method to detect bone peaks, the highest peak from a certain window in the signal is annotated and determined as the bone peak using past experience. Here we used the expert knowledge from a previous study \cite{niu2018situ}. The window position $p_{win}$ and window width $w_{win}$ was shown in TABLE \ref{traditional_parameters}. As the interval $l_{int}$ was already known in Equation (\ref{gt_location}), the start and end position of the window (represented as the unit index) was determined using the following Equations.

\begin{align}\label{win_index}
idx_{start} = (p_{win} - w_{win}/2)/l_{int} \\
idx_{end} = (p_{win} + w_{win}/2)/l_{int}
\end{align}

\begin{table}[t!]
\caption{Window width and position determined by the expert}
\centering
\begin{tabular}{||>{\centering\arraybackslash}m{2cm}|>{\centering\arraybackslash}m{1.5cm}|>{\centering\arraybackslash}m{1.5cm}|>{\centering\arraybackslash}m{1.5cm}||}
\hline
\textbf{LOCATION} & \textbf{CHANNEL} & \textbf{WINDOW POSITION} & \textbf{WINDOW WIDTH} \\
\hline
\hline
\multirow{2}{*}{Femur Epicondyle} & 11 & 12.427mm & 8mm \\
 & 12 & 15.879mm & 9mm \\
\hline
\multirow{3}{*}{Tibia Epicondyle} & 16 & 9.205mm & 7mm \\
 & 17 & 7.939mm & 5mm \\
 & 19 & 12.287mm & 7mm \\
\hline
\multirow{2}{*}{Mid Tibia} & 24 & 3.011mm & 5mm \\
 & 26 & 6.559mm & 5mm \\
\hline
\multirow{2}{*}{Ankle} & 28 & 7.642mm & 6mm \\
 & 29 & 5.211mm & 5mm \\
\hline
\end{tabular}
\label{traditional_parameters}
\end{table}

After cropping the meaningful region, a simple peak detection algorithm from the python library $(scipy.signal.find\_peaks)$ was used. It detected the peak that has a certain height range, prominence, and threshold. In our experiment, the height range was determined using the same training dataset, and tested using the same testing dataset of neural network. The result was shown in the left gray background of TABLE \ref{results_comparison}. Noticed that the bias between the prediction and the ground truth was between 1 to 3 mm, showing that the error was large and the accuracy (below 0.48mm) was quite low.

\begin{table*}[t!]
\centering
\caption{Results comparison with traditional method \\ (left gray part was from traditional method, right bright part was from our CasAtt-UNet.)}
\begin{tabular}{||c|c||>{\columncolor{mygray}}c|>{\columncolor{mygray}}c|>{\columncolor{mygray}}c||c|c|c||}
\hline
\textbf{LOCATION} & \textbf{CHANNEL} & \textbf{ACCURACY} & \textbf{BIAS} & \textbf{FIND PEAK\%} & \textbf{ACCURACY} & \textbf{BIAS} & \textbf{FIND PEAK\%}\\
\hline
\hline
\multirow{2}{*}{Femur Epicondyle} & 11 & 38.30\% & 1.455mm & 92.20\% & 81.18\% & 0.348mm & 91.18\% \\
&12 & 27.84\% & 2.334mm & 86.27\% & 90.07\% & 0.337mm & 92.16\% \\
\hline
\multirow{3}{*}{Tibia Epicondyle} & 16 & 14.21\% & 2.778mm & 100.00\% & 80.22\% & 0.437mm & 92.62\% \\
& 17 & 4.41\% & 2.808mm & 100.00\% & 63.59\% & 0.561mm & 95.59\% \\
& 19 & 19.80\% & 2.488mm & 96.57\% & 81.35\% & 0.275mm & 94.53\% \\
\hline
\multirow{2}{*}{Mid Tibia} & 24 & 32.35\% & 1.302mm & 100.00\% & 66.51\% & 0.554mm & 94.77\%  \\
& 26 & 26.96\% & 1.380mm & 100.00\% & 63.52\% & 0.648mm & 89.53\%  \\
\hline
\multirow{2}{*}{Ankle} & 28 & 28.43\% & 1.294mm & 100.00\% & 55.25\% & 1.616mm & 85.72\%  \\
 & 29 & 20.59\% & 2.211mm & 100.00\% & 59.06\% & 0.623mm & 89.77\% \\
\hline
\end{tabular}
\label{results_comparison}
\end{table*}

\subsection*{B. Deep Learning Method Result}
The result of CasAtt-UNet was shown in the bright right background of TABLE \ref{results_comparison}. Noticed that without any expert knowledge, our model could automatically locate bone peak locations in different areas within sub-millimeter accuracy (except for Channel 28). The average accuracy (below 0.48mm) in all the channels could perform 71.19\% on average, which demonstrated the advantage of our method. It was evident that the traditional method could almost achieve 100\% peak recognition rate. This is because the traditional method could always find a peak in the window region, as long as all samples in the test dataset existed peaks. Thus, it did not provide useful indication of the performance. However for the CasAtt-UNet, it indeed showed a sensitivity issue, as different static probability threshold could produce various length of possible bone peak segments, directly influencing final peak prediction. A better method would be studied later to automatically adjust the probability threshold to achieve the optimal peak recognition performance.

\section{DISCUSSION}

This work constructed a deep-learning-based method for bone peak detection in 1D US signals, to measure the bone positions with A-mode ultrasound transducers when applied in the robotic orthopedic application and exoskeletons. The contextual information preservation and feature localization capability of U-Net were greatly improved by cascading different perception resolution U-Nets together, connected by a novel sampling-based proposal mechanism. The introduced attention blocks enabled the learnt patterns more robust and adapt to more channels situations. 

one limitation of the work is that we only used one cadaver specimen to collect the dataset, the result may be lack of the variability in patient's anatomy, and had not considered other impacts of the surgical environment factors that can directly impact model's performance. Another limitation is that, compared with other bone registration and reconstruction study, our work only provided and evaluated the distance between skin and bones without completing the whole registration process. However, it is worth noting that the advantage of our method lies in the high precision and automatic bone position measurement under skin, without additional trauma or expert knowledge. Even in the registration process, the previous study \cite{niu2018feasibility} reported the registration error as 2.81 mm, which was much worse than our measurement of bone positions.

Besides, this technique could further provide real-time bone locations when installed on the surgical robot arms, as the normal time to process 2D ultrasound images has been removed. Its sub-millimeter accuracy can guide the robot to do high-precision alignment of total knee replacement, also in other similar surgery that had strict requirement. In addition, since the neural network can achieve high precision in identifying the special and sparse bone peaks, this work convinced that deep learning technique was capable to identify and find profiles for very special and irregular signal peaks in the 1D raw signal, which could also inspire other works that require identification of the non-evident and sparse peaks in the 1D raw signal.

\section{CONCLUSIONS}
In this study, we proposed a novel deep-learning structure to detect highly random and sparse bone peaks in the 1D raw signals with high precision. The measured distance can be used for later bone registration and bone position recovery in real-time, which could be installed on the robotics arms and guide the surgery with sub-millimeter accuracy. The experiment results demonstrated the optimal precision we can achieve, which shows the promising prospective of our system to apply in not only the total knee replacement arthroplasty but also other similar robotics surgeries.

\addtolength{\textheight}{-12cm}   % This command serves to balance the column lengths
                                  % on the last page of the document manually. It shortens
                                  % the textheight of the last page by a suitable amount.
                                  % This command does not take effect until the next page
                                  % so it should come on the page before the last. Make
                                  % sure that you do not shorten the textheight too much.

%%%%%%%%%%%%%%%%%%%%%%%%%%%%%%%%%%%%%%%%%%%%%%%%%%%%%%%%%%%%%%%%%%%%%%%%%%%%%%%%

%%%%%%%%%%%%%%%%%%%%%%%%%%%%%%%%%%%%%%%%%%%%%%%%%%%%%%%%%%%%%%%%%%%%%%%%%%%%%%%%

%%%%%%%%%%%%%%%%%%%%%%%%%%%%%%%%%%%%%%%%%%%%%%%%%%%%%%%%%%%%%%%%%%%%%%%%%%%%%%%%

\bibliographystyle{main}
\bibliography{main}

\end{document}